# A Quantized Representation of Intertemporal Choice in the Brain

James Tee, *Member, IEEE*, and Desmond P. Taylor, *Life Fellow, IEEE*

***Abstract*—Value [4][5] is typically modeled using a continuous representation (i.e., a Real number). A discrete representation of value has recently been postulated [6]. A quantized representation of probability in the brain was also posited and supported by experimental data [7]. Value and probability are inter-related via Prospect Theory [4][5]. In this paper, we hypothesize that intertemporal choices may also be quantized. For example, people may treat (or discount) 16 days indifferently to 17 days. To test this, we analyzed an intertemporal task by using 2 novel models: quantized hyperbolic discounting, and quantized exponential discounting. Our work here is a re-examination of the behavioral data previously collected for an fMRI study [8]. Both quantized hyperbolic and quantized exponential models were compared using AIC and BIC tests. We found that 13/20 participants were best fit to the quantized exponential model, while the remaining 7/20 were best fit to the quantized hyperbolic model. Overall, 15/20 participants were best fit to models with a 5-bit precision (i.e., $2^5$ = 32 steps). In conclusion, regardless of hyperbolic or exponential, quantized versions of these models are better fit to the experimental data than their continuous forms. We finally outline some potential applications of our findings.**

## I. Introduction

Intertemporal choice, also known as discounting, focuses on value decision trade-offs at different points in time. For example, would you prefer to receive a $10 payment today (present option), or wait for now and receive a $15 payment next month (future option)? Experimental data on intertemporal choices are typically modeled using either a hyperbolic discounting function [9],[10],[11] or an exponential discounting function [11],[12],[13]. The primary difference between the 2 lies in the steepness of the discounting curves; the hyperbolic function decays at a steeper pace than the exponential discounting function, signifying a value decision preference for the present option, as opposed to a future option. Equivalently, preference for the present option signifies a decision maker who will choose the future option only if the payment amount for the future option is significantly larger (say, $20) than the present option (say, $10). To date, discounting functions have been modeled in terms of continuous Real numbers.

In [14], the authors investigated the question of whether information in the brain is represented in continuous or discrete form. This question is relevant to our work here, because the form of information representation determines which model is best for data analysis. It is worth re-emphasizing here that both the above models (i.e., the hyperbolic discounting and the exponential discounting functions) are historically based on a continuous representation (i.e., Real numbers). By incorporating communication theory drawn from communications systems engineering (e.g., [15],[16]) and Shannon information theory [17], they [14] concluded that information representation in the brain cannot be continuous, due to the presence of noise – but must be represented in a discrete manner. This is a major paradigm shift from traditional approaches to data analysis and modeling of the brain.

In [7], the authors utilized the conclusions drawn from [14] to develop a quantized (i.e., discrete) model of human perception of probability. They compared the continuous model of probability representation with the quantized model, and found that the discrete model is a better fit to experimental data. The findings further reaffirm the hypothesis that information in the brain is represented in a discrete manner. Consistent with the approach outlined in [7], a quantized representation of value was also recently proposed [6].

In this paper, we hypothesize that intertemporal choices (value with a time dimension) are also quantized. For example, people may treat (or discount) 16 days indifferently to 17 days. We re-analyze the experimental data from Cox and Kable [8] using novel quantized (discrete) discounting models, and compare them with conventional, continuous discounting models. The performance of both models was further compared using the Akaike Information Criterion (AIC) and the Bayesian Information Criterion (BIC), arriving at a conclusive quantized (i.e., discrete) result.

Manuscript accepted on September 15, 2020. This work was supported by the National Institutes of Health under Grant EY019889. This paper was presented in part at the Annual Meeting of the Society for Neuroeconomics, Miami, FL, USA, Sept. 2015 [1]. Parts of this paper were also first published in James Tee's Ph.D. dissertation, undertaken and successfully completed at the Department of Psychology of New York University [2]. An earlier version of this paper was also published as an arXiv preprint [3].

James Tee was previously with the Department of Psychology, New York University. He is now with the Communications Research Group, Department of Electrical & Computer Engineering, University of Canterbury, Private Bag 4800, Christchurch 8020, New Zealand (email: james.tee@canterbury.ac.nz).

Desmond P. Taylor is with the Communications Research Group, Department of Electrical & Computer Engineering, University of Canterbury, Private Bag 4800, Christchurch 8020, New Zealand (email: desmond.taylor@canterbury.ac.nz).

## II. A Quantized (Discrete) Hyperbolic Discounting Model

Intertemporal choices are typically modeled using a continuous hyperbolic discounting function of the form [9]:

$$SV = \frac{A}{1+kD}$$
$$\frac{SV}{A} = \frac{1}{1+kD}$$

where $A$ is the objective value, $D$ is the time delay (in units of days), $k$ is the discount rate and $SV$ is the subjective value. Fig. 1 (left) shows an example of a conventional, continuous hyperbolic discounting function.

We quantize [18] the continuous hyperbolic function, resulting in the form:

$$Q_n\left[\frac{SV}{A}\right] = Q_n\left[\frac{1}{1+kD}\right]$$

where $n$ is the number of bits and $Q_n[]$ denotes a quantization function that divides the hyperbolic discounting function into $2^n$ possible steps or quantization levels. Fig. 1 (right) shows an example of a 3-bit quantized hyperbolic discounting model (3 bits = $2^3$ = 8 levels). We note that the conventional, continuous model is simply a quantized model with an infinite number of steps (i.e., quantization levels).

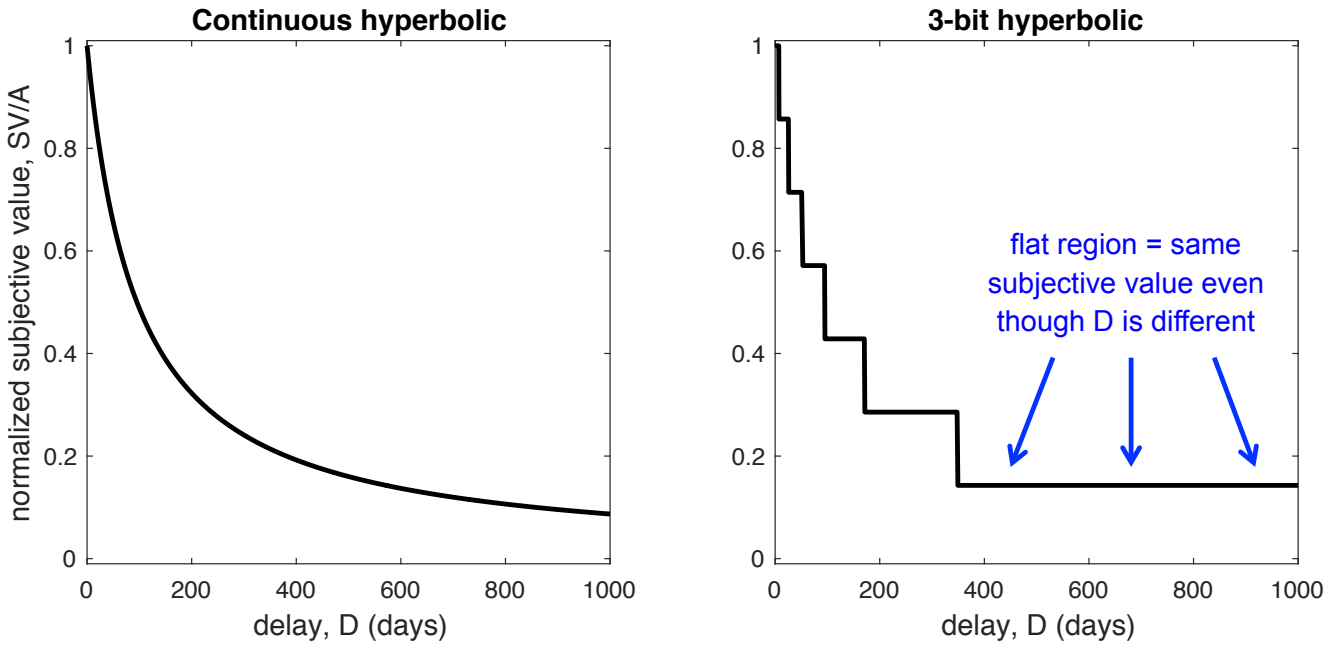

Fig. 1. (Left) Conventional, continuous hyperbolic discounting model. (Right) 3-bit quantized hyperbolic discounting model.

A pertinent question that is worth posing at this stage is whether or not the flat regions (i.e., areas of indifference) in Fig. 1 are possible and plausible. These areas imply the circumstance that v($X,t) = v($X,t+dt) where dt is positive and v() is the subjective value function. We outline below a few examples that support such a circumstance.

Suppose that a financially impoverished John were to borrow $40 from his affluent friend Tom, and John promised Tom that the money would be repaid in 2 months' time. As the deadline approaches, Tom doesn't care if the money were to be paid on day 58 or day 60 (because he isn't in dire need of the $40), as long as the money is repaid as promised. This is a possible and plausible example of v($40,58) = v($40,60), signifying the existence of an area of indifference with t = [58,60]. As the time horizon is stretched further out into the future, the size of this area could (or would) plausibly widen. For example, if John were to promise Tom that the $40 would be repaid in 1 year's time, Tom does not care if the money were to be paid on day 350 or day 365. Likewise, if the time horizon is shortened, it is plausible that this area of indifference could (or would) narrow. For example, if John were to promise Tom that the $40 would be repaid in 1 day's time, Tom is unlikely to care if the money were to be paid at the 23-hour mark or the 24-hour mark. A quantized discounting function allows for the possibility of such areas of indifference, whereas a continuous discounting function is unable to do so.

There are also possible and plausible circumstances under which one would choose the later option (i.e., receiving $X at time t+dt) as opposed to the nearer option (i.e., receiving $X at time t). For example, for personal reasons, Peter may opt for his $40 wage to be paid next week (i.e., t=7), as opposed to tomorrow (i.e., t=1). One such personal reason could arise from Peter's fear that the $40 at hand may be stolen or lost before next week, which is when he needs the $40 to pay his mobile phone bill. Another plausible personal reason would be the lack of self-discipline to not spend the money prematurely. For example, if Peter has $40 in hand, he is more likely to spend the money if he were to walk past a comic book store; consequently, he could deliberately choose to have his wages paid next week so that he would use it to pay for his mobile phone bill instead of spending it beforehand on a comic book. Another possible reason is that Peter is a recovering gambling addict; he knows if he were to have the money in hand, he is more likely to gamble and lose it, and consequently, he chooses to defer having money in hand before he needs it. These behaviors can be accommodated by a quantized discounting function, whereas a continuous discounting function is unable to do so.

Another way to scrutinize this notion of areas of indifference is to explore the plausibility of choosing to receive less money in the future: v($X,t) = v($X-$Y,t+dt), where $Y>0. The logic here is that, if it is plausible to choose to receive less money sometime in the future, then, it would most certainly be plausible to choose to receive an equal amount of money sometime in the future. That is, would it be plausible that one would choose to receive $39 next week, versus receiving $40 today? At first thought, such a scenario would be normatively unthinkable. Yet, in the hypothetical example above of Peter as a recovering gambling addict, he would conceivably make this choice – better to have $39 next week than to not have anything at all next week. Outside the realm of hypothetical examples and experimental research laboratories, similar scenarios have actually taken place in real life. For example, on May 21, 2020, the yields (i.e., interest rates) of the 5-year UK government treasury bonds fell below zero for the first time, with the 2-year yields dropping to -0.062% [19]. In such a situation of negative interest rates, if one were to invest $1,000 today, one would get less than $1,000 back in 2 years' time (i.e., choosing to receive less money in the future). On July 31, 2020, the real yields of the 10-year US Treasury bonds fell to -1% [20]. In literature on intertemporal choice, negative discounting rate behaviors have previously been found, but they remained largely classified as puzzling anomalies [21][22]. While a quantized discounting

function may not be able to fully account for the behaviors exhibited by negative discounting rates, it does allow for the notion of indifferences of value in time, something which a continuous discounting function is unable to do.

## III. Methods

Our analysis is a re-examination of the human behavioral data previously collected for an fMRI study by Cox and Kable [8]. We begin here with a brief outline of their methods, which were approved by the Institutional Review Board of the University of Pennsylvania [8]. During each trial, a participant chooses between 2 options: \$40 now, or \$X in D days (see Fig. 2).

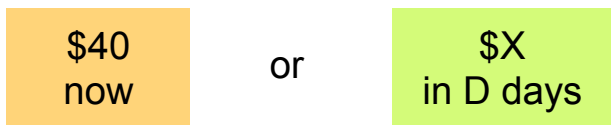

Fig. 2. Stimuli for intertemporal choice experiment.

\$X ranges from \$41 to \$100, while D ranges from 1 to 327 days. There were 204 trials in total. Participants were paid \$20 for their participation in the study. At the end of the experiment, one of the completed 204 trials was randomly selected and a bonus corresponding to the participant's choice in the selected trial was paid. For example, if the randomly selected trial was a choice between receiving \$40 now (present option) versus receiving \$60 in 18 days' time (future option) and the participant had (during the experiment) chosen the present option, then, a \$40 bonus was paid to the participant. If the participant had chosen the future option instead, then, a \$60 bonus was paid to the participant after an 18-day delay. The bonus was paid using a debit card with the corresponding delay date. A total of 20 participants performed the task. In terms of data analysis, we extended the same maximum likelihood estimation approach [23] for data fitting (as in [8]) using the quantized hyperbolic discounting model. We also employed nested hypothesis testing [24], similar to the approach in [7].

## IV. Results for the Quantized Hyperbolic Model

### A. *Fitting Experimental Data*

Consistent with the approach by Cox and Kable [8], we fit the experimental data using logistic regression. Fig. 3 shows the negative log likelihood of the maximum likelihood estimation process for a sample participant. Precision ranged from 1 to 16 bits. The fit for the continuous model is shown in the horizontal dashed blue line. As the quantized precision (i.e., blue line) increases from 1 to 5 bits, the fit improves (i.e., negative log likelihood decreases). Beyond that, the fit becomes worse (i.e., value of negative log likelihood increases) and subsequently flattens off at (i.e., converges to) the same level as the continuous model (i.e., horizontal dashed blue line). For this sample participant, the best fit occurs at a precision of 5 bits, suggesting that a quantized model is a better fit than a continuous model.

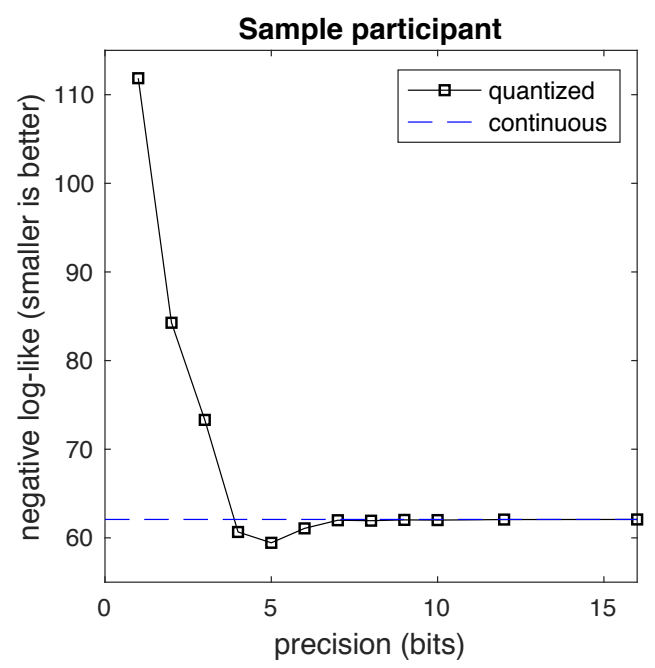

Fig. 3. Negative log likelihood of hyperbolic model fit for one sample participant.

Fig. 4 shows the difference in the log likelihood (LL) between the quantized and continuous hyperbolic models for each of the 20 participants. It can be observed that the magnitude of improvements varies among the participants. However, the quantized hyperbolic model offers an improvement over the continuous hyperbolic model for all 20 participants.

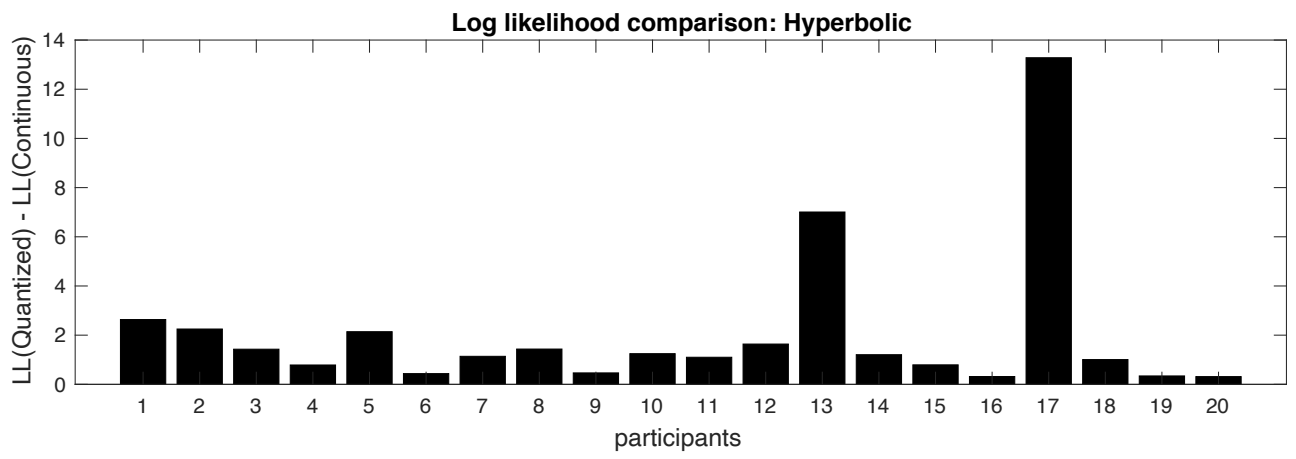

Fig. 4. Difference in log likelihood between the quantized and continuous hyperbolic models for each participant.

Of the 20 participants, 9 were best fit to 5-bit quantized hyperbolic models (i.e., $2^5 = 32$ steps). The histograms of fitted model parameters (i.e., the number of bits of quantization, the noise of the fitting process, and the discounting rate) are shown in Fig. 5. The noise parameter is also known as the beta parameter (i.e., slope) of the logistic function. We note that the largest number of bits resulting from the data fitting exercise is 9 bits, representing a model with $2^9 = 512$ levels; a continuous hyperbolic discounting model is the case of an infinite number of levels.

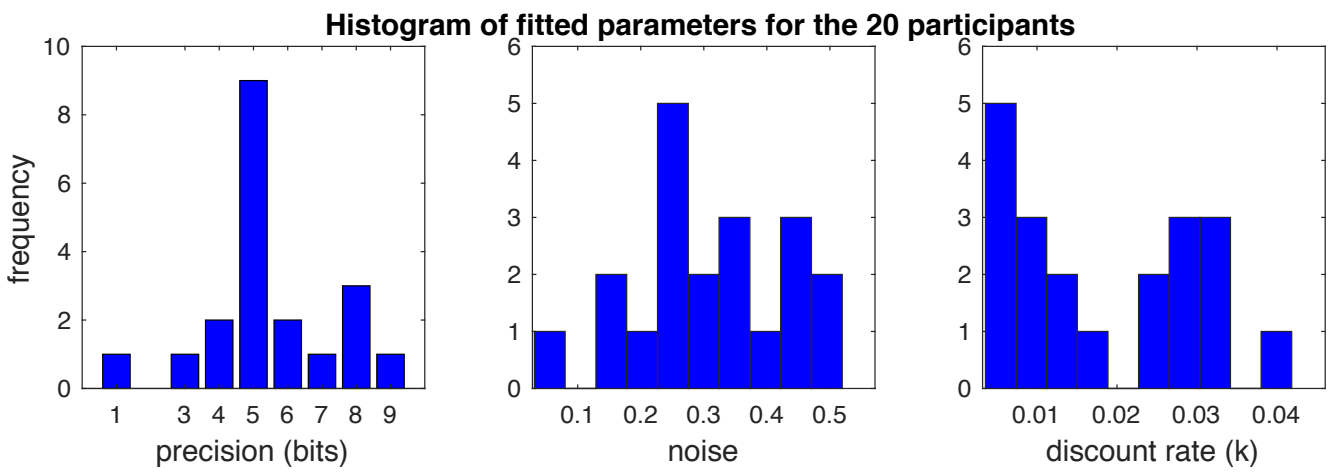

Fig. 5. Histogram of fitted parameters (20 participants) for the quantized hyperbolic model.

### B. *Gauging Robustness and Sensitivity*

In order to gauge the robustness and sensitivity of the 5-bit result in Fig. 5, we further fit only a subset of the experimental data to the quantized hyperbolic model. The logic and rationale were to check whether or not the 5-bit result (i.e., mode) would still hold if some data points were omitted from the data fitting process. If the results were to change drastically, then it would imply that the 5-bit mode (of Fig. 5) was very sensitive to the 204 data points, and therefore, lacking robustness.

We employed an approach that is similar to cross-validation (CV) in machine learning. Note that, we are using the word "similar" here (as opposed "same"), because in machine learning, the cross-validation approach (of systematically leaving out data) is used for training purposes, after which the trained neural network is used for prediction onto a new dataset. Here, we are leaving out data points in order to gauge the sensitivity and robustness of the data fitting process. In our CV approach here, we omitted a window of $m$ consecutive data points (i.e., trials), upon which the subset of data was fit to the quantized hyperbolic model. We ran such a subset/partial data fit multiple times, each time omitting a different window of $m$ consecutive data points. In order to ensure objectivity in our selection/choice of the window, we ran all possible combinations of $m$ consecutive data points. This process was performed for all participants. The cumulative results are plotted in Fig. 6. The left graph, titled CV0, is the result from Fig. 5 (i.e., with the full dataset). The middle graph, titled CV10, is the histogram for $m = 10$ (i.e., obtained from omitting 10 consecutive data points). Despite the data omissions, this CV10 result retains a mode of 5 bits which is consistent with the CV0 result. We repeated the same omission process for $m = 20$ (i.e., omitting 20 consecutive data points), and the result is shown on the right graph, titled CV20. This CV20 result retains a mode of 5 bits which is consistent with both the results of CV0 and CV10. We note here that CV20 signifies an approximately 10% omission of data (i.e., 20 out of 204). Despite this sizeable omission, the 5-bit mode was obtained. The consistency among the CV0, CV10 and CV20 results provide a positive indication that the 5-bit mode of Fig. 5 was robust and insensitive, instead of being a consequence of coincidence or accidental artifacts.

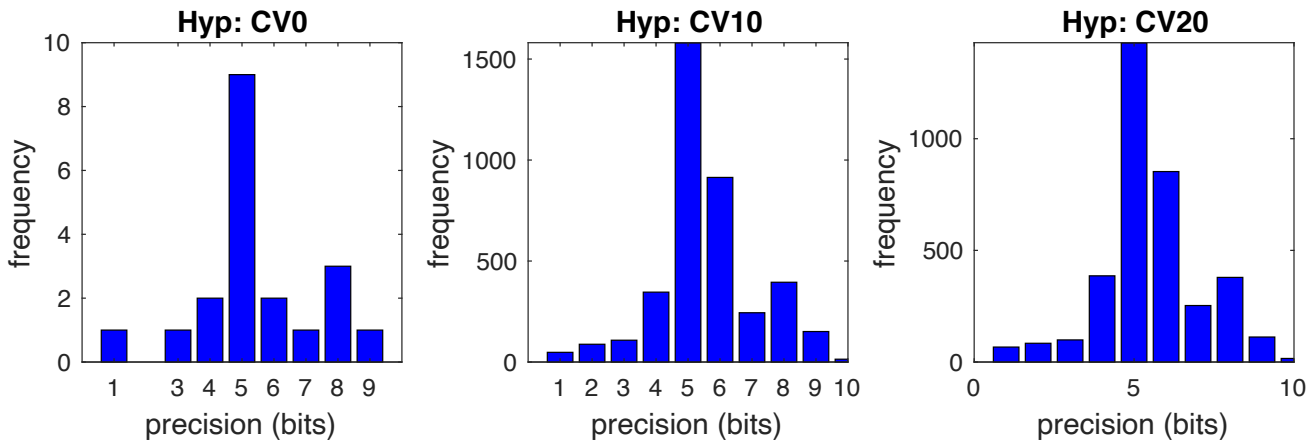

Fig. 6. Histograms arising from fitting a subset of the experimental data to the quantized hyperbolic model.

### C. Bootstrap Simulations to Check for Confound

We next considered whether the 5-bit quantized result could have been confounded with a continuous model – meaning, is it possible that our human participants made choices using a continuous (i.e., 20-bit) model, but our experimental and data fitting processes somehow mistakenly produced 5-bit results? This concern is illustrated in Fig. 7(a). Note that, we used a 20-bit model for convenience, and we reasonably assumed that 20 bits of precision is indistinguishable from a continuous model. From hereon, the term "20-bit" is used interchangeably with "continuous".

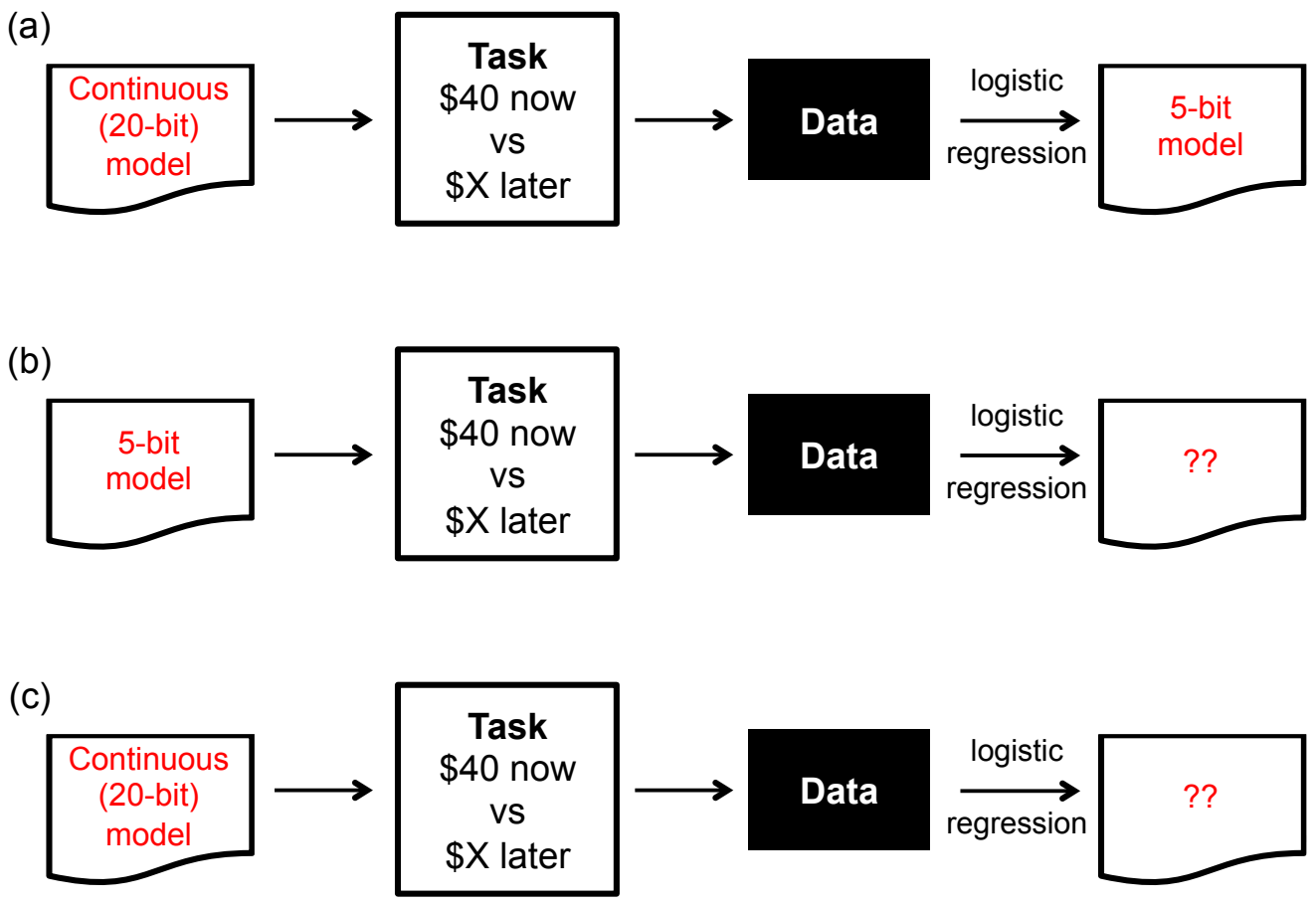

Fig. 7. (a) Could our 5-bit results have been confounded with a continuous decision-maker? (b) First set of bootstrap simulations on a 5-bit decision-maker. (c) Second set of bootstrap simulations on a 20-bit decision-maker.

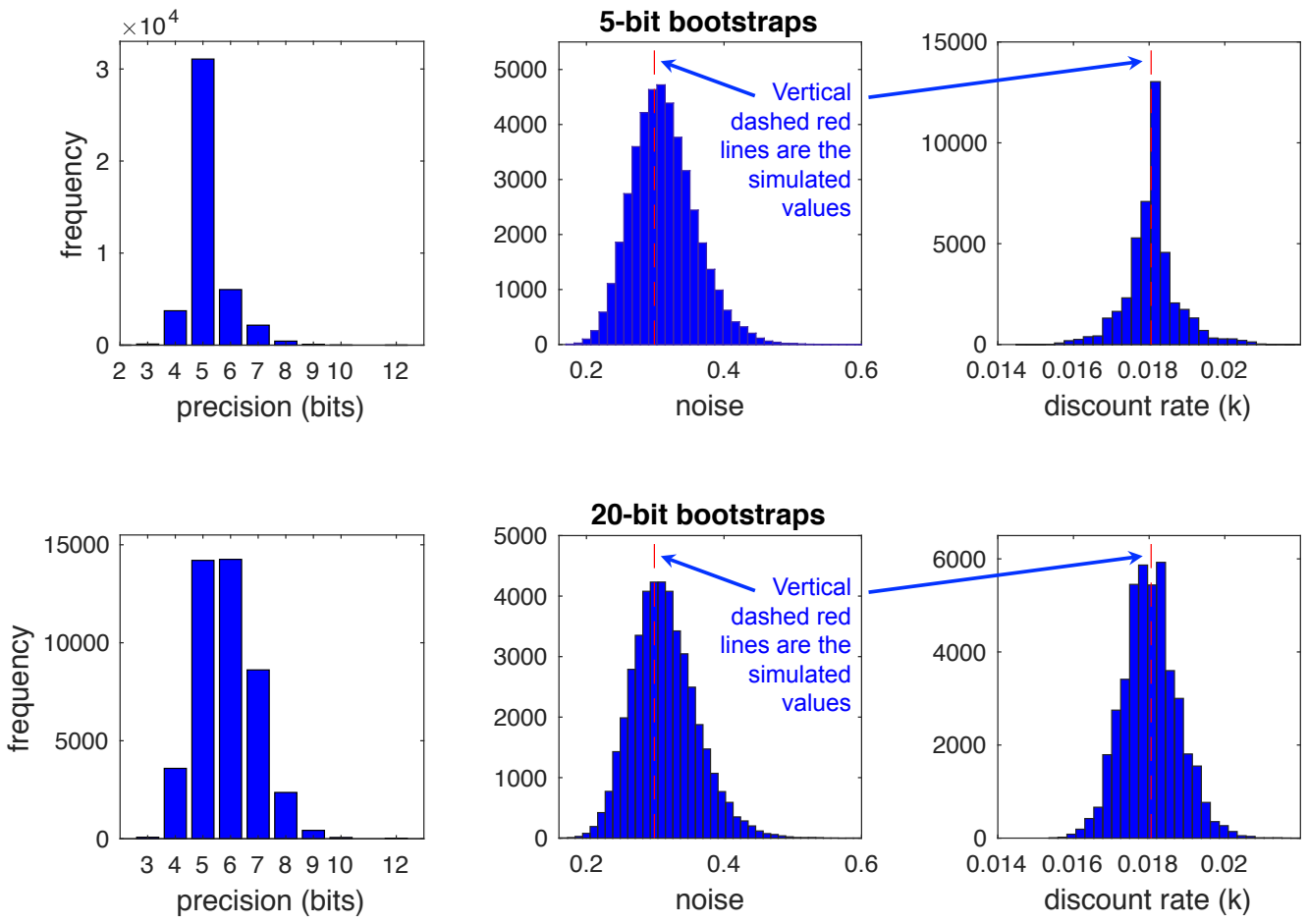

Fig. 8. Results of bootstrap simulations of the 5-bit and 20-bit hyperbolic models.

In order to check for this confound, we performed 2 sets of bootstrap simulations. In the first set of bootstraps, we simulated a 5-bit model as the decision-maker in performing the task, and we looked for the model that was the best fit to this simulated data (see Fig. 7(b)). In the second set of bootstraps, we simulated a continuous (20-bit) model as the decision-maker in performing the task, and we looked for the model that was the best fit to this simulated data (see Fig. 7(c)). Results of both bootstraps are shown in Fig. 8. Results for the 5-bit bootstraps are plotted in row 1, whereas results of the 20-bit bootstraps are plotted in row 2. Column 1 is the precision (in bits), column 2 is the beta value from the logistic regression, and column 3 is the discount rate. In columns 2 and 3, the vertical dashed red lines represent the simulated values (i.e., the values used as the decision-maker in performing the task) and we see that the histograms flank the simulated values as we expected. The key parameter is the precision (column 1). For the 5-bit bootstraps, we see that the mode of the histogram is 5 bits, as expected. On the other hand, for the 20-bit bootstraps, we see that there are 2 modes in the histogram, at 5 bits and 6 bits. When we visually compare the histograms of the bootstrap simulation (Fig. 8, column 1) with the one from our actual experimental data (in Fig. 5, left plot), we can see that the 5-bit single mode histogram is more consistent, as opposed to the bi-modal histogram from the 20-bit bootstraps. This provides a positive indication that our 5-bit experimental result is unlikely to be confounded with a 20-bit (continuous) model.

### *D. Statistical Tests for Confound*

Instead of simply relying on the visual positive indication, we performed 2 further tests to compare whether the distribution of the experimental data (i.e., Fig. 5, left plot) is statistically similar to the null hypothesis distribution (i.e., 20-bit bootstraps of Fig. 8, bottom left). Note that we were unable to use the standard Kolmogorov-Smirnov (K-S) Goodness-of-Fit test here because the K-S test only applies to continuous distributions and the distribution must be fully specified instead of being estimated from the data [25]. In our case, the distribution of the experimental data is hypothesized to be discrete (i.e., quantized) and the null hypothesis distribution is obtained via bootstrap simulations (i.e., estimated instead of specified). First, we performed the standard Chi-square test:

$$\chi^2 = \sum_i \frac{(O_i - E_i)^2}{E_i}$$

where $O$ is the observed frequency and $E$ is the expected frequency. The null hypothesis was rejected at $p < 0.0001$. Secondly, we performed a G-test [26]:

$$G = 2\sum_i O_i \log\left(\frac{O_i}{E_i}\right)$$

The null hypothesis was rejected at $p < 0.001$. Given that both statistical tests rejected the null hypothesis (i.e., 20-bit model), we are as certain as we can be that the 5-bit quantized result obtained from our experimental data is highly unlikely to be confounded with a continuous model.

### *E. Nested Hypothesis Tests*

Our quantized hyperbolic discounting model has 2 free parameters (i.e., $n$ and $k$). Since the experimental data has a mode of 5 bits, we applied a nested hypothesis test [24] [7] on the model with precision fixed at 5 bits instead of being a free parameter:

$$Q_n\left[\frac{1}{1+kD}\right] \quad \rightarrow \quad Q_5\left[\frac{1}{1+kD}\right]$$

The model on the left has 2 free parameters (i.e., $n$ and $k$) whereas the model on the right has only 1 free parameter (i.e., $k$). The purpose of the nested hypothesis test [24] [7] is to explore whether the second parameter is statistically justifiable or required for the data fitting of each participant. We note that such a 1-parameter ($k$ only) model is analogous to the conventional, continuous hyperbolic discounting model [9] except that $n$ is fixed at 5 bits instead of being fixed at infinity. Results of the nested hypothesis test showed that 17 out of 20 participants were best fit to this 1-parameter model ($k$ is a free parameter while $n$ is fixed at 5 bits). The 5-bit quantized hyperbolic discounting curves for two representative participants are shown in Fig. 9.

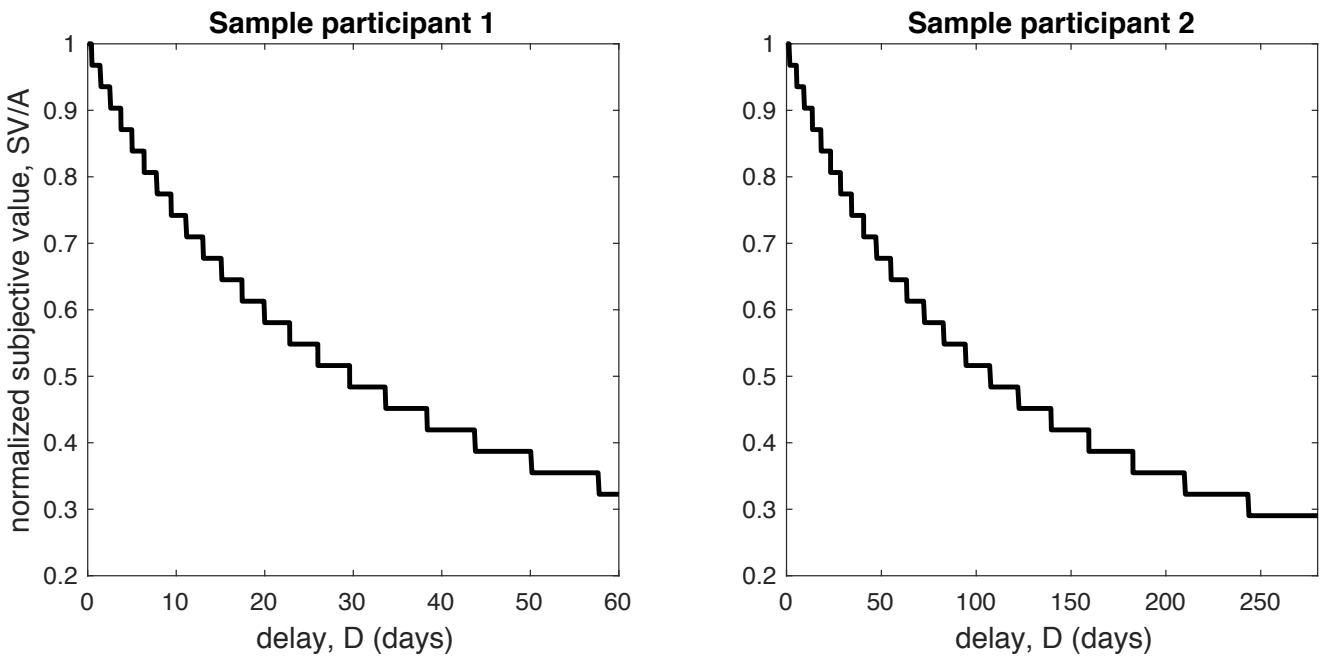

Fig. 9. 5-bit quantized hyperbolic discounting curves for two representative participants.

## V. A Quantized (Discrete) Exponential Discounting Model

Another commonly used discount function is the continuous exponential discounting model [12] [13]:

$$\frac{SV}{A} = \delta^D$$

where $SV$ is the subjective value, $A$ is the objective value, $D$ is the time delay, and $\delta$ is the discount rate with $0 < \delta < 1$. Fig. 10 (left) shows an example of a conventional, continuous exponential discounting function. We note that in some literature [11], the exponential discounting model is expressed as:

$$\frac{SV}{A} = e^{-bD}$$

where $b$ is the discount rate parameter. In our work here, we adopted the mathematically equivalent version [12] [13], where:

$$\frac{SV}{A} = e^{-bD} = \left(e^{-b}\right)^D = \left(\frac{1}{e^b}\right)^D = \delta^D$$

Similar to the hyperbolic case, we quantized [18] this model to produce:

$$Q_n\left[\frac{SV}{A}\right] = Q_n\left[\delta^D\right]$$

where $2^n$ is the number of steps. Fig. 10 (right) shows an example of a 3-bit quantized exponential discounting model (i.e., 3 bits = $2^3$ = 8 levels). Similar to the case of a quantized hyperbolic model, the continuous exponential model is simply a quantized model with an infinite number of steps.

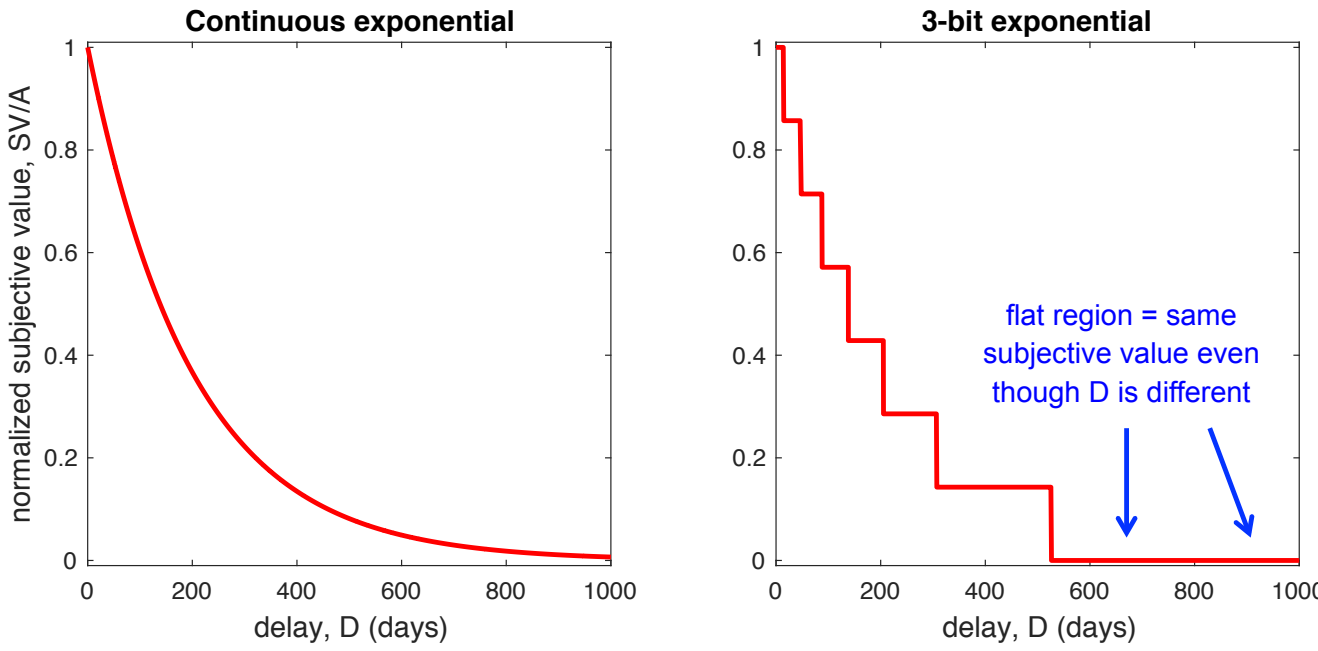

Fig. 10. (Left) Conventional, continuous exponential discounting model. (Right) 3-bit quantized exponential discounting model.

## VI. Results for the Quantized Exponential Model

### A. Fitting Experimental Data

We fit the same experimental data to the quantized exponential model using the same maximum likelihood estimation method [8]. Fig. 11 shows the negative log likelihood for a sample participant. Similar to the case of the quantized hyperbolic model, the precisions for the quantized exponential model here range from 1 to 16 bits. The fit for the continuous exponential model is shown in the horizontal dashed blue line. As the precision of the quantized exponential model (i.e., blue line) increases from 1 to 5 bits, the fit improves (i.e., value of negative log likelihood decreases). Beyond that, the fit becomes worse (value of negative log likelihood increases) and subsequently flattens off at (i.e., converges to) the same level as the continuous model. For this sample participant, the best fit occurs at a precision of 5 bits, suggesting that a quantized model is a better fit than a continuous one, similar to that observed in the hyperbolic case.

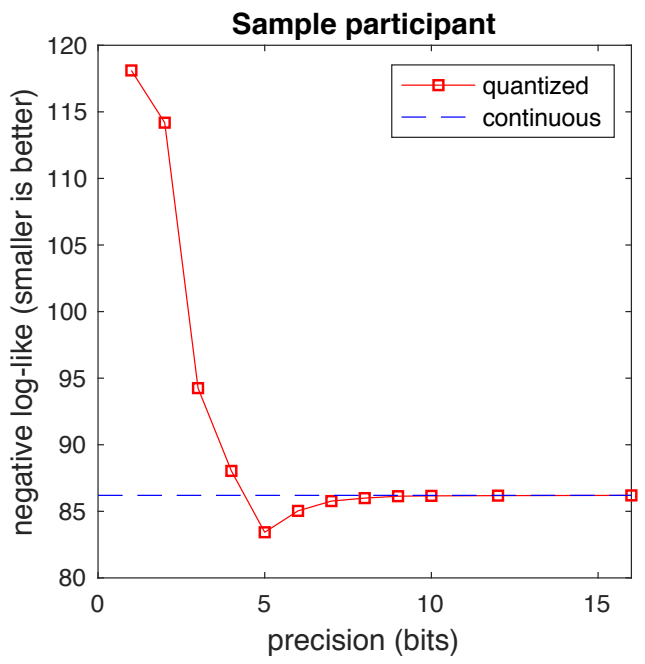

Fig. 11. Negative log likelihood of exponential model fit for one sample participant.

Fig. 12 shows the difference in the log likelihood (LL) between the quantized and continuous exponential models for each of the 20 participants. While the magnitude of improvements varies among the participants, the quantized exponential model offers an improvement over the continuous exponential model for all 20 participants. We found that 8 out of 20 participants were best fit to 5-bit quantized exponential models (i.e., $2^5 = 32$ steps). The histograms of fitted parameters (i.e., the number of bits of quantization, the noise of the fitting process, and the discounting rate) are shown in Fig. 13.

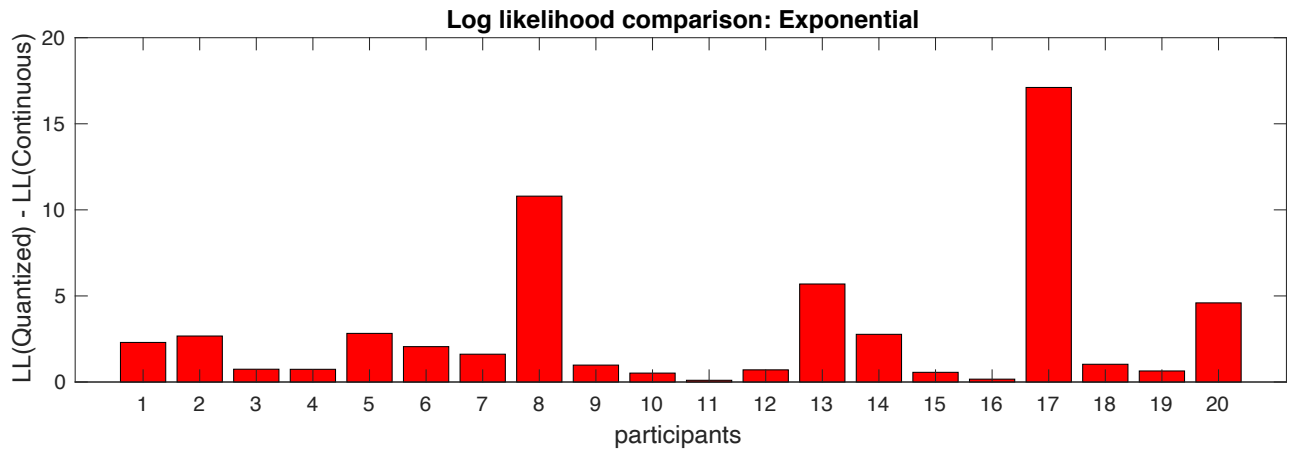

Fig. 12. Difference in log likelihood between the quantized and continuous exponential models for each participant.

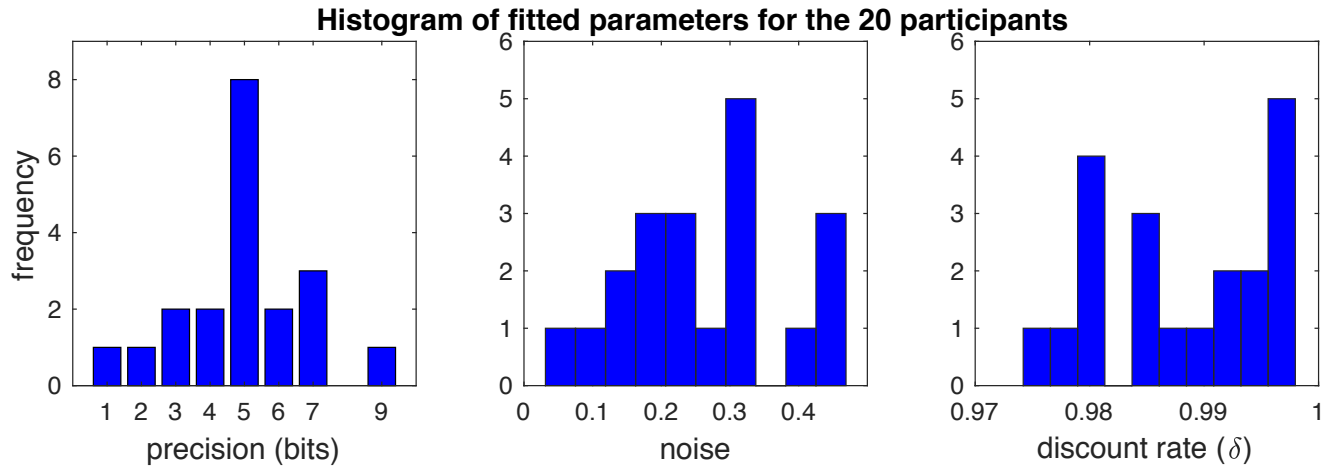

Fig. 13. Histogram of fitted parameters (20 participants) for the quantized exponential model.

### B. Gauging Robustness and Sensitivity

In order to gauge the robustness and sensitivity of the 5-bit result in Fig. 13, we further fit only a subset of the experimental data to the quantized exponential model, employing the same approach as described in Section IV-B for the quantized hyperbolic model. The cumulative results are plotted in Fig. 14. The left graph, titled CV0, is the result from Fig. 13 (i.e., with the full dataset). The middle graph, titled CV10, is the histogram obtained from omitting 10 consecutive data points. The right graph, titled CV20, is the histogram obtained from omitting 20 consecutive data points. Both the CV10 and CV20 results retained a mode of 5 bits, which is consistent with the CV0 result. The consistency among the CV0, CV10 and CV20 results provide a positive indication that the 5-bit mode of Fig. 13 was robust and insensitive, instead of being a consequence of coincidence or accidental artifacts.

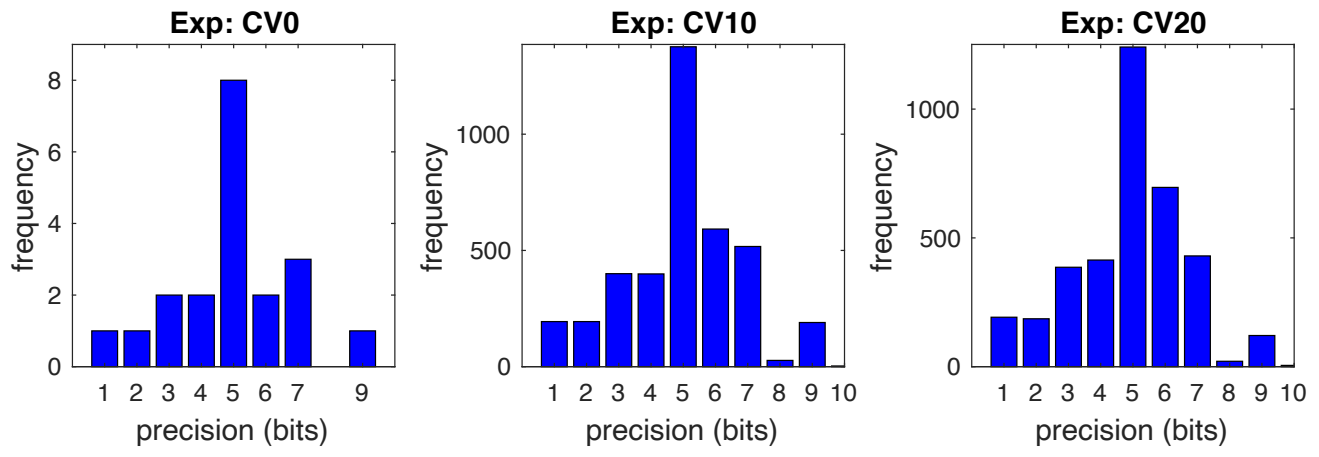

Fig. 14. Histograms arising from fitting a subset of the experimental data to the quantized exponential model.

### C. Bootstrap Simulations to Check for Confound

We next examined whether our 5-bit quantized result could have been confounded with a continuous model – that, perhaps our human participants made choices using a continuous (i.e., 20-bit) exponential model, but our experimental and data fitting processes somehow mistakenly produced 5-bit results? Similar to how we tested the quantized hyperbolic model, we performed 2 sets of bootstrap simulations – one using a 5-bit exponential model as the decision-maker in performing the task (see Fig. 7(b)), and one using a 20-bit exponential model (i.e., equivalent to and indistinguishable from a continuous model) as the decision-maker (see Fig. 7(c)) – and looked for the model that was the best fit to these simulated data. Results of both bootstraps are shown in Fig. 15. The 5-bit bootstraps are plotted in the top row, whereas the 20-bit bootstraps are plotted in the bottom row. Column 1 is the precision (in bits), column 2 is the beta value from the logistic regression, and column 3 is the discount rate. In columns 2 and 3, the vertical dashed red lines represent the simulated values and we see that the histograms flank the simulated values as we expected. The key parameter that is of primary interest is the precision (column 1). For the 5-bit bootstraps, we see that the mode of the histogram is clearly 5 bits, as expected. For the 20-bit bootstraps, we see that there are almost 2 modes in the histogram, at 5 bits and 6 bits. A visual comparison of the precision of 5-bit bootstrap (Fig. 15, top left) with the experimental data (in Fig. 13, left plot) gives us a positive indication that our 5-bit experimental result is unlikely to be confounded with a 20-bit (continuous) model (Fig. 15, bottom left).

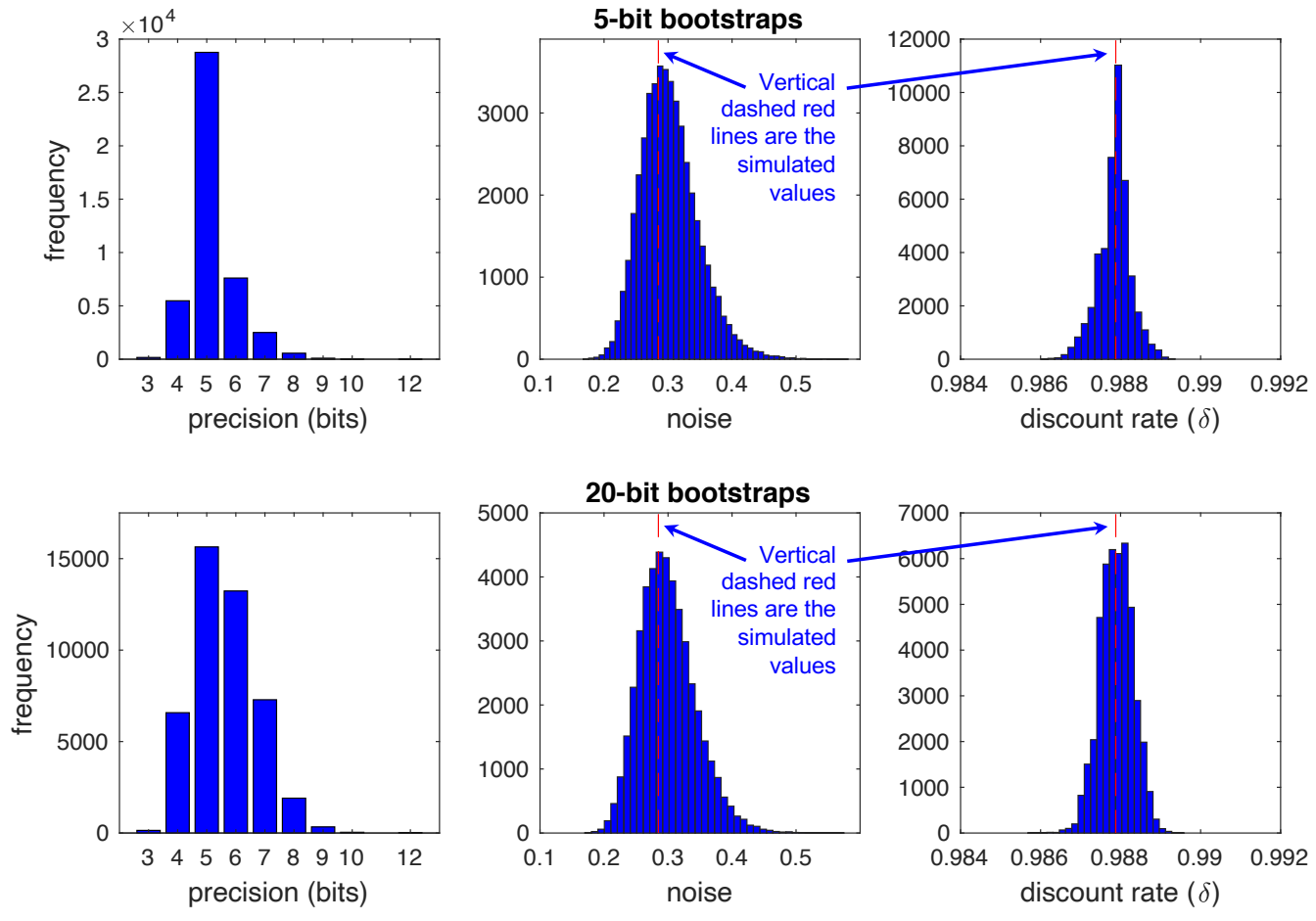

Fig. 15. Results of bootstrap simulations of the 5-bit and 20-bit exponential models.

## D. Statistical Tests for Confound

Similar to the quantized hyperbolic case, we performed 2 further tests to compare whether the distribution of our experimental data (i.e., Fig. 13, left plot) is statistically similar to the null hypothesis distribution (i.e., 20-bit bootstraps of Fig. 15, bottom left). For the Chi-square test, the null hypothesis was rejected at $p < 0.0001$. For the G-test, the null hypothesis was rejected at $p < 0.01$. Given that both statistical tests rejected the null hypothesis (i.e., 20-bit model), we are as certain as we can be that the 5-bit quantized result obtained from our experimental data is highly unlikely to be confounded with a continuous model.

## E. Nested Hypothesis Tests

As was with the case for the quantized hyperbolic model, we applied a nested hypothesis test to explore whether the second parameter is statistically justifiable or required for the data fitting of each participant:

$$Q_n\left[\delta^D\right] \quad \rightarrow \quad Q_5\left[\delta^D\right]$$

The quantized exponential model on the left has 2 free parameters (i.e., $n$ and $\delta$), whereas the model on the right has only 1 free parameter (i.e., $\delta$) with $n$ being fixed at 5 bits (instead of being a second free parameter). The results from the nested hypothesis test showed that 15 out of 20 participants were best fit to this 1-paramater model (i.e., $\delta$ is a free parameter while $n$ is fixed at 5 bits). The 5-bit quantized exponential discounting curves for two representative participants are shown in Fig. 16.

To summarize, 8 out of the 20 participants were best fit to the 5-bit quantized exponential models. After applying the nested hypothesis test, 15 out of the 20 participants were best fit to the 1-parameter quantized exponential model. These exponential findings are consistent with the hyperbolic ones.

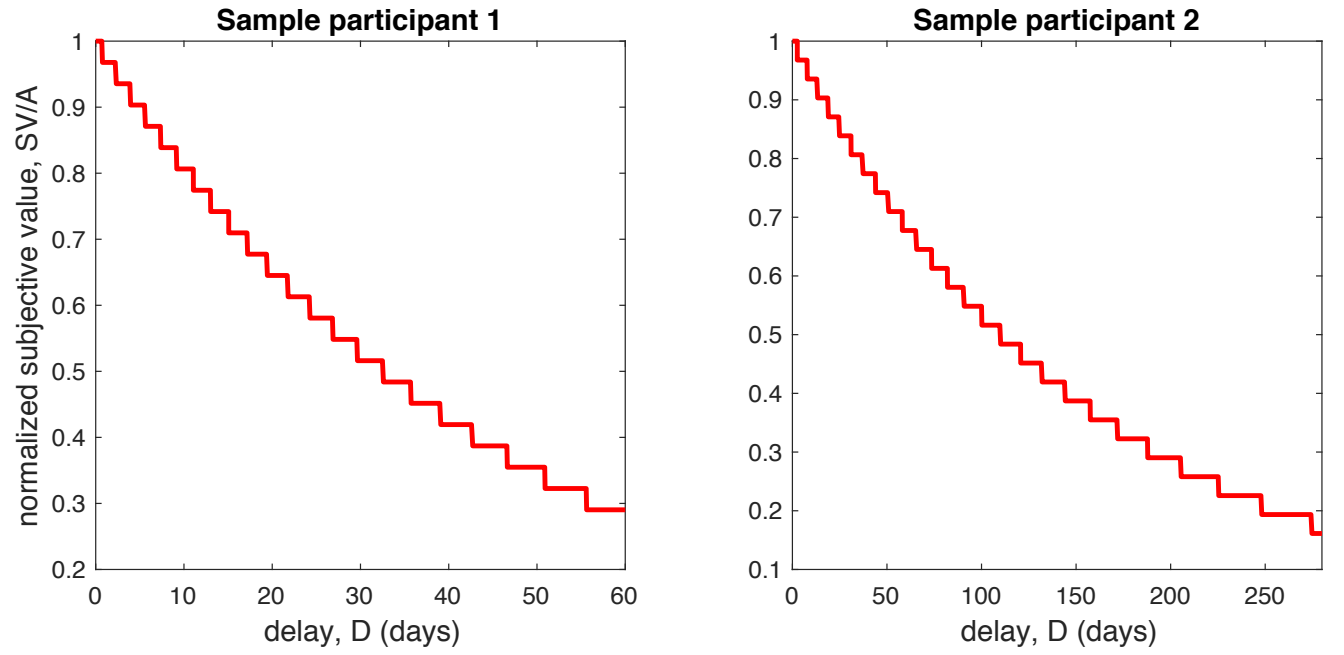

Fig. 16. 5-bit quantized exponential discounting curves for two representative participants.

## VII. Comparing the Quantized Hyperbolic and Quantized Exponential Models

While one participant's data may be an excellent fit to a quantized hyperbolic model, it is possible that the same participant's data may be an even better fit to a quantized exponential model. The reverse can also be true for a different participant. In other words, some participants' data may be more suited to be modeled by the quantized hyperbolic model, whereas other participants' data may be more suited to be modelled by the quantized exponential model. For completeness, we compared the performance of the quantized hyperbolic model with the quantized exponential model. We took the best fit quantized hyperbolic models (i.e., after the nested hypothesis test) and compared it with the best fit quantized exponential models (after the nested hypothesis test) using the Akaike Information Criterion (AIC) and the Bayesian Information Criterion (BIC) (see [27] for an overview of AIC and BIC). These comparisons are plotted in Fig. 17. Note that, for both the AIC and BIC comparisons, a smaller value represents a better fit. Both AIC and BIC results are in agreement: 13 out of 20 participants were best fit to the quantized exponential model, with the remaining 7 participants best fit to the quantized hyperbolic model. Following this best-of-the-best AIC/BIC comparison, 15 out of 20 participants have 5-bit precision (see Fig. 18). A comparison of the quantized exponential and quantized hyperbolic curves of 2 representative participants is shown in Fig. 19.

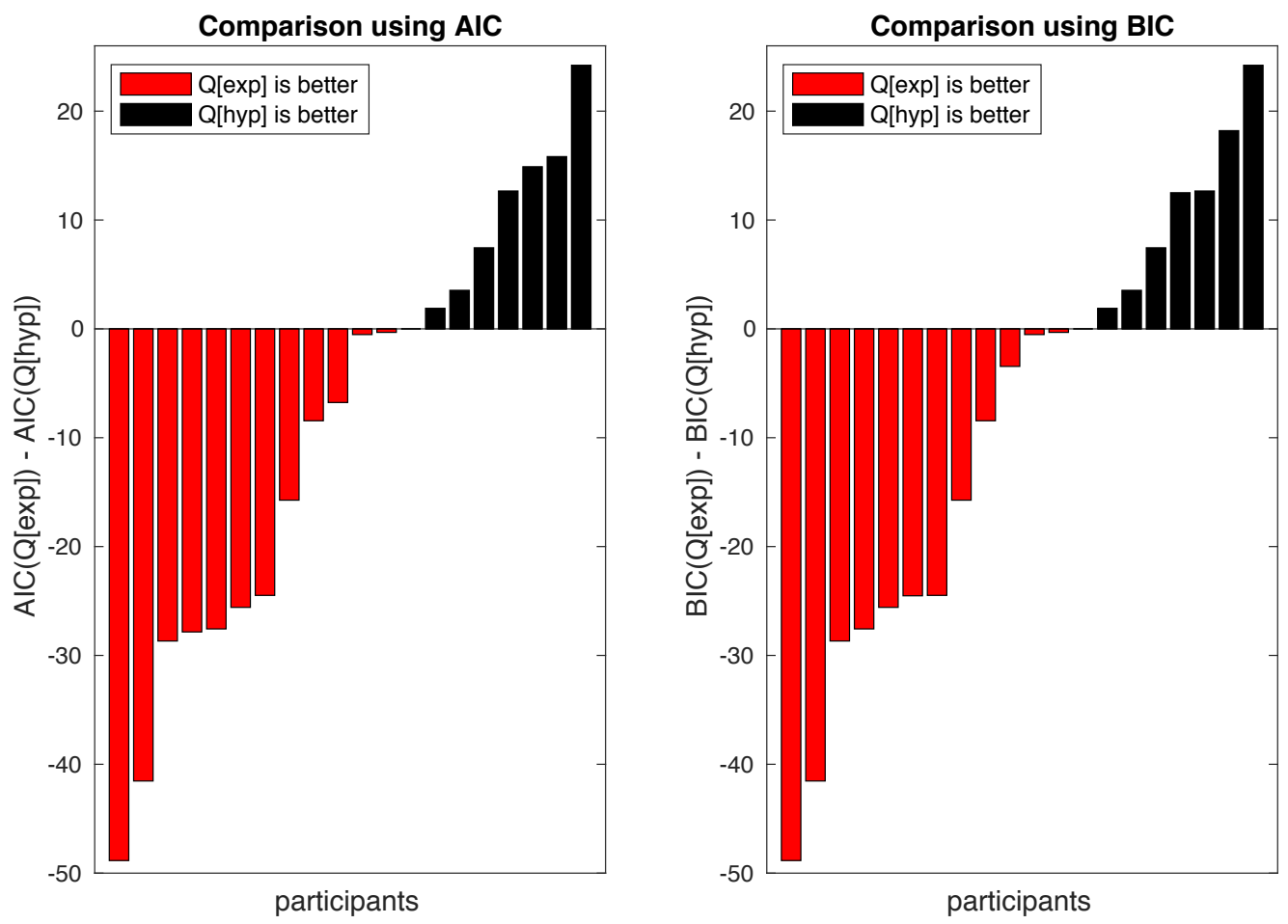

Fig. 17. Comparison using the Akaike Information Criterion (AIC) (left) and Bayesian Information Criterion (BIC) (right).

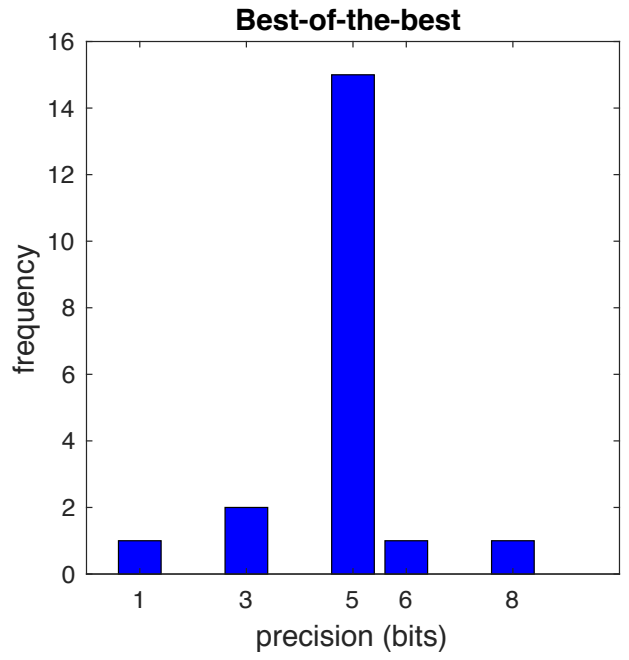

Fig. 18. Histogram of the 20 participants' precisions after the best-of-the-best AIC/BIC comparison.

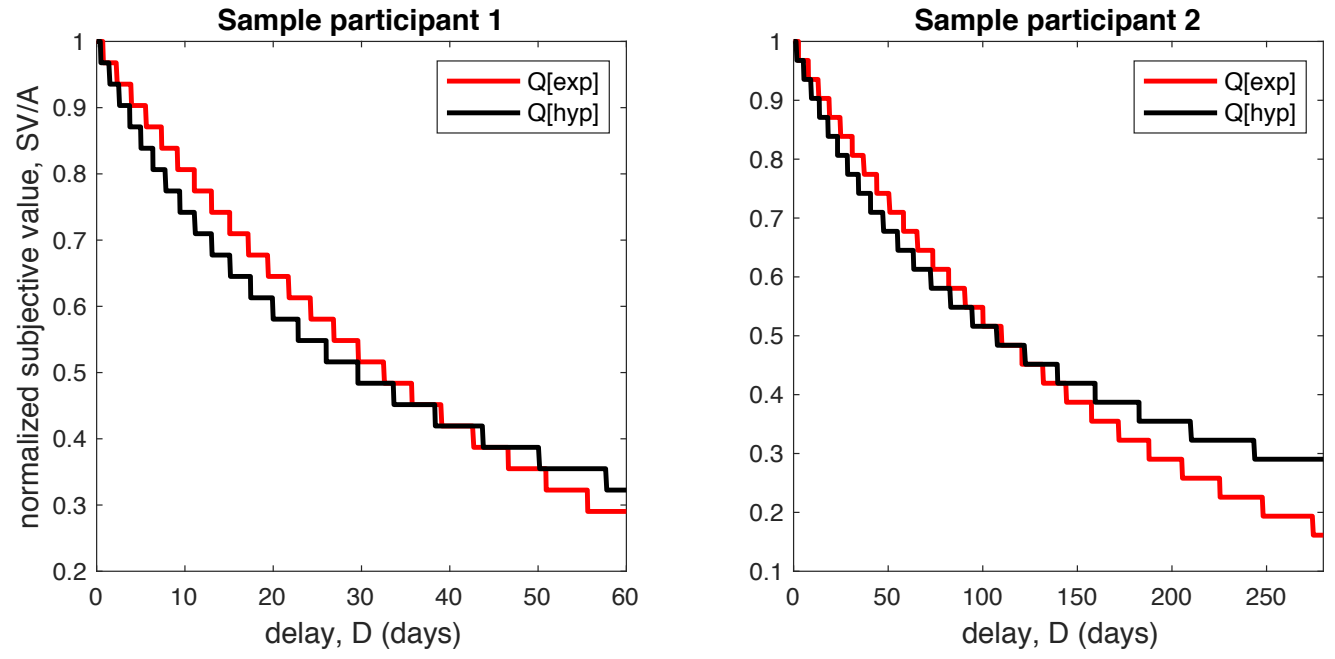

Fig. 19. Comparing the quantized exponential and quantized hyperbolic curves of 2 representative participants.

## VIII. Conclusions

In summary, we reiterate that 13/20 participants were best fit to a quantized exponential model while the remaining 7 are best fit to a quantized hyperbolic model. Overall, 15/20 (i.e., 75%) participants were best fit to models exhibiting 5-bit precision. The most important conclusion is that, regardless of whether we are using a hyperbolic or an exponential discounting model, their quantized versions are a better fit to the experimental data than their respective continuous versions. These results confirmed our intuitive hypothesis – that, humans categorize (or chunk) time. Our results here also reaffirm the discrete conclusions reported in [14] [7]. While continuous models have, up till now, been convenient for analyzing experimental data, we should be open to the real possibility that actual decisions are quantized (i.e., discrete). Given that our quantized result here was obtained based on an independent study (a study that was neither designed nor conducted by us), we are confident that our approach is generalizable to many existing and future studies.

One relevant application of our findings is in understanding debt-related (i.e., spend-now-pay-later) decisions (e.g., credit cards, loans, mortgages) [28]. Another relevant application is in studying health-related choices (e.g., ignore the broccoli, enjoy the fried chicken now, and face the health/cholesterol consequences later) [29]. Our findings are also relevant to clinical and behavioral research on addictions (e.g., alcohol, drugs) [30].

## Acknowledgment

J. Tee thanks C. R. Gallistel, M. Woodford and J. W. Kable for their helpful comments and suggestions. J. Tee further thanks J. W. Kable for sharing his experimental data. J. Tee and D. P. Taylor thank the late W. H. Tranter for his constructive comments and suggestions.

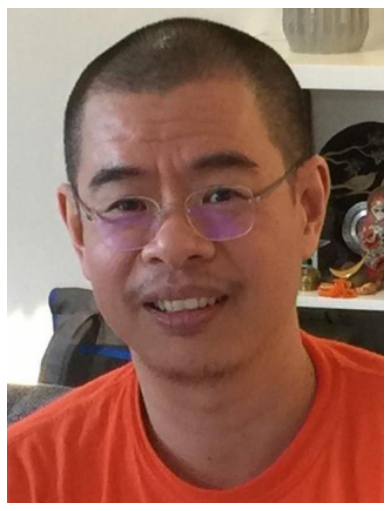

James Tee (M'17) completed his Ph.D. in Electrical & Electronic Engineering at the University of Canterbury in 2001, where he worked on Turbo Codes under the supervision of Des Taylor. Subsequently, he held various industry and policy positions at Vodafone Group, the World Economic Forum, New Zealand's Ministry of Agriculture & Forestry, and the United Nations. To facilitate his career transitions, he pursued numerous supplementary trainings, including an MBA at the Henley Business School, and an MPhil in Economics (Environmental) at the University of Waikato. In 2012, James began his transition into scientific research at New York University (NYU), during which he completed an MA in Psychology (Cognition & Perception) and a PhD in Experimental Psychology (Neuroeconomics). Afterwards, he worked as an Adjunct Assistant Professor at NYU's Department of Psychology, and a Research Scientist (Cognitive Neuroscience) at Quantized Mind LLC. Since 2017, he is an Adjunct Research Fellow at the University of Canterbury. James is an Eastern medicine physician, with an MS in Acupuncture from Pacific College of Oriental Medicine. In August 2020, he began further training in Substance Abuse Counseling at the New School for Social Research (NSSR). His current research interests in neuroscience focuses on reverse engineering the communications codebook (i.e., signal constellation) of the Purkinje cell neuron. James is also working on Artificial Intelligence (AI) approaches inspired by insights drawn from psychology (cognition, perception, decision-making) and neuroscience.

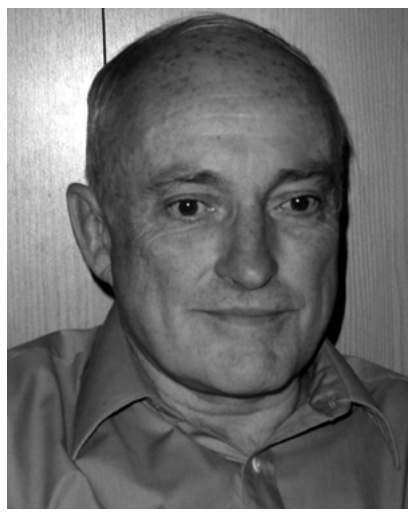

Desmond P. Taylor (LF'06) received the Ph.D. degree in electrical engineering from McMaster University, Hamilton, ON, Canada, in 1972. From 1972 to 1992, he was with the Communications Research Laboratory and the Department of Electrical Engineering, McMaster University. In 1992, he joined the University of Canterbury, Christchurch, New Zealand, as the Tait Professor of communications. He has authored approximately 250 published papers and holds several patents in spread spectrum and ultra-wideband radio systems. His research is centered on digital wireless communications systems focused on robust, bandwidth-efficient modulation and coding techniques, and the development of iterative algorithms for joint equalization and decoding on fading, and dispersive channels. Secondary interests include problems in synchronization, multiple access, and networking. He is a Fellow of the Royal Society of New Zealand, the Engineering Institute of Canada, and the Institute of Professional Engineers of New Zealand.